\documentclass{article}

\begin{document}

\title{Laboratory Density Functionals}

\author{B. G. Giraud \\
bertrand.giraud@cea.fr, Service de Physique Th\'eorique, \\
DSM, CE Saclay, F-91191 Gif/Yvette, France}

\date{\today} 
\maketitle

\begin{abstract}

We compare several definitions of the density of a self-bound system, such 
as a nucleus, in relation with its center-of-mass zero-point motion. A 
trivial deconvolution relates the internal density to the density defined
in the laboratory frame. This result is useful for the practical definition
of density functionals.

\end{abstract}

\bigskip
Recently \cite{Eng}, Engel pointed out that the usual form of nuclear 
Hamiltonians,
\begin{equation}
H=\sum_{i=1}^A \frac{p_i^2} {2m} + \sum_{i>j=1}^A v_{ij}\, ,
\label{hamil0}
\end{equation}
forbids a proper definition of the nuclear density. Indeed, the center
of mass (CM) of the system delocalizes in a plane wave; the nucleus is 
everywhere, the density is flat. The conclusion holds whether the two-body 
interaction $v$ is local or non local and still holds if three-body forces 
are also present in $H,$ as long as any explicit density dependence is 
avoided; Galilean invariance must be ensured.

An elementary modification of $H,$
\begin{equation}
{\cal H}=H + A\, \omega^2\, R^2/2,\ \ \ \ R=A^{-1} \sum_i r_i\, ,
\label{hamiloc}
\end{equation}
traps the CM. The ground state of ${\cal H}$ now factorizes
as a product of a Gaussian for this CM and an ``internal'' wave 
function of the $(A-1)$ Jacobi coordinates,
$$
\Psi(r_1,r_2,...,r_A)=\Gamma(R)\ \psi_{int}(\xi_1,\xi_2,...,\xi_{A-1})\, ,
$$
\begin{equation}
\xi_1=r_2-r_1,\,
\xi_2=r_3-\frac{r_2+r_1}{2},\, ...\, ,
\xi_{A-1}=r_A-\frac{r_{A-1}+r_{A-2}+...+r_1}{A-1}\, .
\label{fctr}
\end{equation}
Calculations in the $\{r_i\}$ representation are much more convenient than 
those in the $\{R,\xi_j\}$ one, for obvious symmetrization reasons. The 
laboratory density,
\begin{equation}
\rho(r)=A \int dr_1\, dr_2\, ...\, dr_{A-1}\  
|\Psi(r_1,r_2,...,r_{A-1},r)|^2\, ,
\label{lab}
\end{equation}
is much easier to calculate than the ``internal'' density,
\begin{equation}
\sigma(\xi)=A \int d\xi_1\, d\xi_2\, ...\, d\xi_{A-2}\  
|\psi_{int}(\xi_1,\xi_2,...,\xi_{A-2},\xi)|^2\, ,
\label{int}
\end{equation}
Throughout this note, we shall use the word ``internal'' instead of 
``intrinsic'' when we refer to properties independent from the CM. This is 
because we retain the word ``intrinsic'' for those states out of which 
rotation bands and/or parity vibrations are modelized. Our understanding is 
that the adjectives ``internal'' and ``intrinsic'' belong to completely 
distinct concepts and models.

Three remarks are in order at this stage,

\noindent
i) The density of interest for a density functional theory (DFT) is $\rho,$ 
not $\sigma.$ Indeed the Hohenberg-Kohn theorem derives from embedding the 
system in an external field, namely replacing ${\cal H}$ by
\begin{equation}
{\cal K}={\cal H} + \sum_{i=1}^A u(r_i)\, ,
\end{equation}
then considering the density $\tau(r)$ of the ground state $\Xi$ of 
${\cal K},$ 
\begin{equation}
\tau(r)=A \int dr_1\, dr_2\, ...\, dr_{A-1}\  
|\Xi(r_1,r_2,...,r_{A-1},r)|^2\, ,
\end{equation}
and finally proving that there is a one-to-one map between $u$ and $\tau.$
The ground state energy of ${\cal K}$ receives the contribution 
$\int dr\, u(r)\, \tau(r),$ out of which $u$ and $\tau$ are recognized as 
conjugate Legendre coordinates, hence the functional Legendre transform which
defines the density functional. When $u$ vanishes, $\rho$ is that limit of 
$\tau$ which minimizes the functional. It seems obviously very difficult 
to set any similar chain of arguments in the Jacobi representation!

ii) Despite this priority of $\rho$ for a DFT, the 
internal nature of $\sigma$ is compelling for a physical interpretation. It is
tempting to calculate $\sigma$ in the $\{r_i\}$ representation. With
the Gaussian $\Gamma$ square normalized to unity, one finds
$$
A^{-1} \sigma(\xi) = \int dR\, d\xi_1\, d\xi_2\, ...\, d\xi_{A-2}\ 
[\Gamma(R)]^2\, |\psi_{int}(\xi_1,\xi_2,...,\xi_{A-2},\xi)|^2 =
$$
$$
\int dR\, d\xi_1 ... d\xi_{A-2} d\xi_{A-1}\ 
\delta(\xi_{A-1}-\xi)\, [\Gamma(R)]^2\,
|\psi_{int}(\xi_1,\xi_2,...,\xi_{A-2},\xi_{A-1})|^2 =
$$
\begin{equation}
\int dr_1 ... dr_{A-1} dr_A\  
\delta\left(r_A-\frac{r_1+r_2+ ... +r_{A-1}}{A-1}-\xi\right)\,
|\Psi(r_1,r_2,...,r_A)|^2.
\label{deltrick} 
\end{equation}

\noindent
iii) the last Jacobi coordinate also reads
\begin{equation}
\xi_{A-1}=\frac{A\, r_A - (r_1+r_2+...+r_{A-1}+r_A)}{A-1} = 
\frac{A}{A-1}\, (r_A-R).
\label{xir}
\end{equation}
The density $\sigma$ therefore also represents, except for a trivial 
rescaling factor, the density referring to the internal degree of freedom, 
$r_A-R.$

Finally, according to Eqs. (\ref{lab}) and (\ref{xir}), the density $\rho$ 
reads,
$$
A^{-1} \rho(r) = \int dr_1\, dr_2\, ...\, dr_{A-1}\, dr_A\  
\delta(r_A-r)\, |\Psi(r_1,r_2,...,r_{A-1},r_A)|^2 =
$$
$$
\int dR\, d\xi_1 ... d\xi_{A-1}\  
\delta\left(R+\frac{A-1}{A}\, \xi_{A-1}-r\right)\, [\Gamma(R)]^2\, 
|\psi_{int}(\xi_1,...,\xi_{A-2},\xi_{A-1})|^2 =
$$
$$
\int dR\, d\xi_1 ... d\xi_{A-2}\  [\Gamma(R)]^2\ \left|
\psi_{int}\left[\xi_1,...,\xi_{A-2},\frac{A}{A-1}\, (r-R)\right]\right|^2 =
$$
\begin{equation}
A^{-1} \int dR\ [\Gamma(R)]^2\ \sigma\left[\frac{A}{A-1}\, (r-R)\right].
\label{convol} 
\end{equation}
The convolution transforming $\sigma$ into $\rho$ is transparent, with again 
an inessential rescaling factor $A/(A-1).$  The zero-point motion of the 
CM blurrs the internal density in a way which can be easily 
inverted, via a deconvolution. The bottom line is, $\rho$ contains the same 
information as $\sigma.$ It is thus possible, and likely much easier, to 
design a DFT with laboratory densities $\tau.$ At the end one recovers the
internal $\sigma$ by a deconvolution of $\rho,$ that solution obtained by 
the minimization, with respect to $\tau,$ of such a ``laboratory 
Hohenberg-Kohn functional''.

It will be noticed that the presence in ${\cal K}$ of external potentials 
$u(r_i)$ couples the CM degree of freedom $R$ and the internal 
ones $\xi_j.$ There is, in general, no CM factorization for 
eigenstates of ${\cal K}.$ The factorization occurs at the limit $u=0.$ 
Then one must verify that the {\it Fourier transform} of $\rho$ shows 
the factorized, Gaussian decay at large momenta, implied by the convolution, 
Eq. (\ref{convol}). Otherwise, deconvolution will fail. This ``deconvolution 
syndrome'' is very well documented in the literature about generator 
coordinates. For the DFT, expansions of the density in 
harmonic oscillator functions and related polynomials, constrained \cite{Gir}
to satisfy matter conservation, make a useful precaution to avoid the
deconvolution syndrome.

For atoms and molecules, CM traps factorizing other wave packets $\Gamma$ 
than Gaussians might be convenient, but the link between $\sigma$ and $\rho$ 
remains the same.

We conclude by claiming that a density functional theory for self-bound 
systems is available with densities in the {\it laboratory} system, without 
any loss of information about the internal, physical density.

\noindent
{\it Acknowledgement}: The author is indebted to B.K. Jennings for calling his
attention to the work of Engel \cite{Eng}.

\end{document}